\documentclass[%
 reprint,
preprintnumbers,
 amsmath,amssymb,
 aps,
floatfix,
]{revtex4-2}
\usepackage{graphicx}% Include figure files
\usepackage{dcolumn}% Align table columns on decimal point
\usepackage{bm}% bold math
\usepackage{color}
\usepackage{physics}
\usepackage{subcaption}

\begin{document}

\preprint{YITP-25-172}

\title{Conformal Bootstrap with Duality-Inspired Fusion Rule}% Force line breaks with \\

\author{Yu Nakayama}
\email{yu.nakayama@yukawa.kyoto-u.ac.jp}
\author{Toshiki Onagi}
\email{toshiki.onagi@yukawa.kyoto-u.ac.jp}
\affiliation{% 
Yukawa Institute for Theoretical Physics, Kyoto University
}%


%\affiliation{
%\affiliation{
%\affiliation{%


\date{\today}% It is always \today, today,

\begin{abstract}
We present a systematic exploration of conformal field theories (CFTs) constrained by duality-inspired fusion rules using the conformal bootstrap. 
We classify the operator spectrum into three sectors: $[\sigma]$, $[\epsilon]$, and $[1]$. The $[\sigma]$ sector consists of all $\mathbb{Z}_{2}$-odd operators. The $\mathbb{Z}_{2}$-even operators are further divided into the $[\epsilon]$ sector, which contains only the operators that change sign under duality, and the $[1]$ sector, which encompasses all remaining operators.
We impose a selection rule motivated by Kramers-Wannier duality, specifically forbidding the appearance of the $[\epsilon]$ sector in the $[\epsilon] \times [\epsilon]$ operator product expansion. By applying this constraint to the lowest-lying relevant scalars, we derive bounds on their conformal dimensions $(\Delta_\sigma, \Delta_\epsilon)$ in dimensions $d=2$ through $d=7$. Our bounds correctly allow the $d=2$ Ising model while excluding the $d=3$ Ising model, demonstrating the effectiveness of the imposed condition. Furthermore, we observe a distinct feature in $d=2$ corresponding to the $\mathcal{M}(8,7)$ minimal model and find non-trivial constraints in $d=3$ ($\Delta_\sigma \gtrsim 0.85$), relevant for theories like QED$_3$. This work opens a new avenue for non-perturbatively probing the landscape of CFTs constrained by fusion rules. 
%\begin{description}
%\end{description}
\end{abstract}

\maketitle

%\tableofcontents

\section{Introduction}
A central challenge in theoretical physics is the fine-tuning problem, prevalent in both high-energy and condensed matter physics. Conventionally, this is resolved by a symmetry principle, leading to ``technical naturalness" \cite{tHooft:1979rat} in high-energy physics or ``self-organized criticality" \cite{Bak:1987xua} in condensed matter physics. The recent development of generalized, or categorical, symmetries—which are often non-invertible—provides a new mechanism to enforce naturalness even where conventional symmetries are absent.

A canonical example is the Kramers-Wannier (KW) duality~\cite{Kramers:1941kn, Kramers:1941zz},  in spin systems or electromagnetic duality in Quantum Electrodynamics (QED). This strong-weak duality non-perturbatively fixes the critical point by identifying it as a self-dual point, thereby solving the fine-tuning problem for the relevant parameter (e.g., temperature) without calculating any quantum or statistical corrections. Such dualities are now understood as instances of non-invertible categorical symmetry~\cite{Frohlich:2004ef, Aasen:2016dop, Bhardwaj:2017xup, Kaidi:2021xfk}. This raises a fundamental question: What is the general class of critical systems protected by this new form of symmetry?

In this paper, we address this question using the conformal bootstrap~\cite{El-Showk:2012cjh, Kos:2014bka, Simmons-Duffin:2015qma, Bobev:2015vsa, Kos:2016ysd, Simmons-Duffin:2016wlq, Dymarsky:2017yzx, Behan:2017rca, Erramilli:2020rlr, Reehorst:2021ykw, Reehorst:2021hmp, Su:2022xnj,He:2023ewx, Chang:2024whx,Simmons-Duffin:2016gjk, Poland:2018epd, Chester:2019wfx}, a rigorous non-perturbative approach to critical phenomena that bounds or predicts critical exponents. While the conformal bootstrap has been extensively applied to systems constrained by conventional (invertible) symmetries \cite{Rattazzi:2010yc,Kos:2013tga,Berkooz:2014yda,Nakayama:2014lva,Nakayama:2014yia,Caracciolo:2014cxa,Nakayama:2014sba,Bae:2014hia,Chester:2014gqa,Kos:2015mba,Chester:2015lej,Li:2016wdp,Nakayama:2017vdd,Iliesiu:2017nrv,Li:2017kck,Stergiou:2018gjj,Kousvos:2018rhl,Stergiou:2019dcv,Kousvos:2019hgc,Reehorst:2019pzi,Chester:2019ifh,Nakayama:2019jvm,Henriksson:2020fqi,Li:2020bnb,He:2020azu,Chester:2020iyt,Reehorst:2020phk,Manenti:2021elk,Ghosh:2021ruh,Baume:2021chx,Sirois:2022vth,Erramilli:2022kgp,Chester:2022hzt,Nakayama:2024iiw,Reehorst:2024vyq,Bartlett-Tisdall:2024mbx,Kousvos:2025ext,Chester:2025uxb}, its power has not yet been leveraged for categorical symmetries (see e.g., \cite{Lin:2023uvm} for a modular bootstrap approach). We present the first conformal bootstrap investigation of critical systems subject to selection rules inspired by duality, providing novel bounds on the space of allowed critical theories.

Concretely, we propose a classification of primary operators into three distinct sectors: the identity sector $[1]$, the spin sector $[\sigma]$, and the energy sector $[\epsilon]$. This classification is motivated by a $\mathbb{Z}_2$ symmetry and a duality-inspired selection rule. The $[\sigma]$ sector consists of all $\mathbb{Z}_{2}$-odd operators. The $\mathbb{Z}_{2}$-even operators are further divided into $[\epsilon]$ and $[1]$ based on their transformation under a hypothetical duality operation: the $[\epsilon]$ sector contains the operators that change sign under the duality, while the $[1]$ sector consists of all other remaining operators. We postulate the following fusion rules among these sectors:
\begin{equation}
\label{eq:ising_fusion_rule}
[\sigma]\times[\sigma] \sim [1]+[\epsilon], \quad [\epsilon]\times[\epsilon] \sim [1],\quad [\sigma]\times[\epsilon] \sim [\sigma].
\end{equation}
The crucial assumption is the absence of $[\epsilon]$ on the right-hand side of the $[\epsilon] \times [\epsilon]$ operator product expansion (OPE). This truncation is the key structure that guarantees the technical naturalness of the lowest-dimension scalar $\epsilon$ in the $[\epsilon]$ sector, as it protects the corresponding parameter from self-correction.

In our bootstrap analysis, we focus on the simplest realization exhibiting this structure. We assume the existence of a single relevant scalar $\sigma$ in the $[\sigma]$ sector, a single relevant scalar $\epsilon$ in the $[\epsilon]$ sector, and no relevant scalars other than the identity operator in the $[1]$ sector.

Let us motivate this fusion rule. In two dimensions, such a structure arises from the non-invertible KW duality line defect~\cite{Kramers:1941kn, Kramers:1941zz, Frohlich:2004ef, Aasen:2016dop, Bhardwaj:2017xup, Kaidi:2021xfk}. The duality defect $D$ maps the order operator sector $[\sigma]$ to the disorder operator sector $[\tilde{\sigma}]$ via $D [\sigma] = [\tilde{\sigma}] D$, and acts non-trivially on the $[\epsilon]$ sector as $D [\epsilon] = -[\epsilon] D$. For OPE compatibility at the self-dual point, the fusion rules Eq.~\eqref{eq:ising_fusion_rule} are required.
While the extension of such 0-form non-invertible symmetries to higher dimensions is non-trivial and typically involves higher-form symmetries, our goal is to explore the landscape of CFTs that phenomenologically satisfy these duality-inspired selection rules for local operators, regardless of the precise nature of the underlying topological defects. To our knowledge, this is the first conformal bootstrap study to impose such a selection rule.

In this paper, we derive bootstrap bounds on critical exponents for CFTs that possess the fusion rules \eqref{eq:ising_fusion_rule}. While our primary motivation is the duality discussed above, this OPE structure may arise from different underlying mechanisms. For instance, consider a CFT with an invertible $U(1)$ symmetry, possibly broken to $\mathbb{Z}_n$ with $n\ge 6$. 
Denoting a complex scalar operator with charge $q \in \mathbb{Z}/2$ as $\Phi_q$, one can construct a realization $\sigma = \Phi_{1/2} + \Phi_{1/2}^\dagger$ and $\epsilon = \Phi_1 + \Phi_1^\dagger$. Charge conservation ensures that $\epsilon$ does not appear in the $\epsilon \times \epsilon$ OPE. For this specific case to be technically natural, we must further assume that all neutral scalar operators $\Phi_0$ beyond the identity are irrelevant.

Nonetheless, such a system, if it exists, must satisfy our bounds. We will later discuss whether such scenarios can be realized in gauge theories in three and four dimensions.

\section{The Conformal Bootstrap with Duality-Inspired Fusion Rules}

We consider a $d$-dimensional unitary CFT with an invertible $\mathbb{Z}_2$ symmetry. We study the minimal system exhibiting this fine-tuning by assuming that the two lowest-lying operators—the $\mathbb{Z}_2$-odd scalar $\sigma$ and the $\mathbb{Z}_2$-even scalar $\epsilon$—are the \textit{only} relevant operators in the theory.

The bootstrap program is implemented by considering the system of four-point functions involving $\sigma$ and $\epsilon$: $\langle \sigma\sigma\sigma\sigma \rangle$, $\langle \epsilon\epsilon\epsilon\epsilon \rangle$, and  $\langle \sigma\sigma\epsilon\epsilon \rangle$.
Applying the OPE constraints of Eq. \eqref{eq:ising_fusion_rule}  to this system, the crossing equations (in terms of cross-ratios $u$ and $v$) are given by:
% \begin{widetext}
    \begin{equation}
        \label{eq:bootstrap_with_kw}
        \begin{aligned}
        0 &= F_{-, \Delta=0, l=0}^{\sigma\sigma,\sigma\sigma}(u,v) + \lambda_{\sigma\sigma \epsilon}^2F_{-, \Delta=\Delta_\epsilon, l=0}^{\sigma\sigma,\sigma\sigma}(u,v) \\
        &\quad +\sum_{\mathcal{O}^+\neq 1, \epsilon}\lambda_{\sigma\sigma \mathcal{O}}^2 F_{-, \Delta, l}^{\sigma\sigma,\sigma\sigma}(u,v),\\
        0 &= F_{-, \Delta=0, l=0}^{\epsilon\epsilon,\epsilon\epsilon}(u,v) +\sum_{\mathcal{O}^+\neq1, \epsilon}\lambda_{\epsilon\epsilon \mathcal{O}}^2 F_{-, \Delta, l}^{\epsilon\epsilon,\epsilon\epsilon}(u,v),\\
        0 &= \lambda_{\sigma\epsilon\sigma}^2F_{-, \Delta=\Delta_\sigma, l=0}^{\sigma\epsilon,\sigma\epsilon}(u,v) + \sum_{\mathcal{O}^-, \mathcal{O}\neq \sigma}\lambda_{\sigma\epsilon \mathcal{O}}^2 F_{-, \Delta, l}^{\sigma\epsilon,\sigma\epsilon}(u,v),\\
        0 &= F_{\mp, \Delta=0, l=0}^{\sigma\sigma,\epsilon\epsilon}(u,v) + \sum_{\mathcal{O}^+\neq1, \epsilon}\lambda_{\sigma\sigma \mathcal{O}}\lambda_{\epsilon\epsilon \mathcal{O}} F_{\mp, \Delta, l}^{\sigma\sigma,\epsilon\epsilon}(u,v)\\ 
        &\quad \pm \lambda_{\sigma\epsilon\sigma}^2 F_{\mp, \Delta=\Delta_\sigma, l=0}^{\epsilon\sigma,\sigma\epsilon}(u,v) 
        \pm \sum_{\mathcal{O}^-\neq \sigma} (-1)^l \lambda_{\sigma\epsilon \mathcal{O}}^2 F_{\mp, \Delta, l}^{\epsilon\sigma,\sigma\epsilon}(u,v),
        \end{aligned}
    \end{equation}
% \end{widetext}
where $\mathcal{O}^+$ runs over the set of $\mathbb{Z}_2$-even operators (which must have even spin), and $\mathcal{O}^-$ runs over the set of $\mathbb{Z}_2$-odd operators (with any spin). Unitarity requires the OPE coefficients $\lambda_{O_1 O_2 O_3}$ to be real.
Here, we have defined the convolved conformal blocks $F^{ij,kl}_{\pm,\Delta,\ell}(u, v)$ as
\begin{equation}
    F^{ij,kl}_{\pm,\Delta,\ell}(u, v) \equiv v^{\frac{\Delta_k+\Delta_j}{2}} g_{\Delta,\ell}^{\Delta_{ij},\Delta_{kl}}(u, v) \pm u^{\frac{\Delta_k+\Delta_j}{2}} g_{\Delta,\ell}^{\Delta_{ij} ,\Delta_{kl}}(v, u). 
\end{equation}
See \cite{Rattazzi:2008pe, Kos:2014bka} for our convention of the conformal block $g_{\Delta,\ell}^{\Delta_{ij},\Delta_{kl}}$.

In Eq.~\eqref{eq:bootstrap_with_kw}, we have explicitly isolated the contributions from the identity operator and the relevant scalars: $\sigma$ in the $\sigma\times\epsilon$ channel and $\epsilon$ in the $\sigma\times\sigma$ channel.  Our duality-inspired constraint $\epsilon \not\in \epsilon \times \epsilon$ is explicitly implemented by the absence of the $\lambda_{\epsilon\epsilon\epsilon}^2$ term in the second line and the $ \lambda_{\sigma \sigma \epsilon}\lambda_{\epsilon\epsilon \epsilon}$ term in the fourth line.

To formulate this system for the semi-definite program (SDP), it is convenient to rewrite Eq.~\eqref{eq:bootstrap_with_kw} in a compact vector notation:
% \begin{widetext}
    \begin{equation}\label{eq:bootstrap_eq}
    \begin{aligned}
        0&=\vec{V}_{\mathrm{identity}} + \lambda_{\sigma\sigma\epsilon}^2 \mqty(  
        F_{-, \Delta=\Delta_{\epsilon}, l=0}^{\sigma\sigma,\sigma\sigma}(u,v)\\
        0\\
        F_{-, \Delta=\Delta_{\sigma}, l=0}^{\sigma\epsilon,\sigma\epsilon}(u,v)\\
        F_{-, \Delta=\Delta_{\sigma}, l=0}^{\epsilon\sigma,\sigma\epsilon}(u,v)\\
        -F_{+,\Delta=\Delta_{\sigma}, l=0}^{\epsilon\sigma,\sigma\epsilon}(u,v)
        )\\
        &\quad + \sum_{\mathcal{O}^{+}\neq 1, \epsilon}
        \begin{pmatrix} \lambda_{\sigma\sigma\mathcal{O}} & \lambda_{\epsilon\epsilon\mathcal{O}} \end{pmatrix}
        \vec{V}_{+,\Delta,\ell}
        \begin{pmatrix} \lambda_{\sigma\sigma\mathcal{O}} \\ \lambda_{\epsilon\epsilon\mathcal{O}} \end{pmatrix}\\
        &\quad +
        \sum_{\mathcal{O}^{-}\neq \sigma}
        \lambda_{\sigma\epsilon\mathcal{O}}^2
        \vec{V}_{-,\Delta,\ell}, 
    \end{aligned}
    \end{equation}
% \end{widetext}
where the vectors $\vec{V}_{+, \Delta, l}$, $\vec{V}_{-, \Delta, l}$, and $\vec{V}_{\mathrm{identity}}$ are given by 
\[
            \vec{V}_{-, \Delta, l} = \mqty(0\\0\\ F_{-, \Delta, l}^{\sigma\epsilon,\sigma\epsilon}(u,v)\\ 
        (-1)^l F_{-, \Delta, l}^{\epsilon\sigma,\sigma\epsilon}(u,v)\\
        -(-1)^l F_{+, \Delta, l}^{\epsilon\sigma,\sigma\epsilon}(u,v)),
\]
\[
      \vec{V}_{+, \Delta, l} 
      = \mqty(
      \mqty(F_{-, \Delta, l}^{\sigma\sigma,\sigma\sigma}(u,v)& 0\\0& 0)\\
      \mqty(0& 0\\0&  F_{-, \Delta, l}^{\epsilon\epsilon,\epsilon\epsilon}(u,v)) \\
      \mqty(0& 0\\0&0) \\
      \mqty(0& \frac{1}{2}F_{-, \Delta, l}^{\sigma\sigma,\epsilon\epsilon}(u,v)\\ \frac{1}{2}F_{-, \Delta, l}^{\sigma\sigma,\epsilon\epsilon}(u,v)& 0)\\
      \mqty(0& \frac{1}{2}F_{+, \Delta, l}^{\sigma\sigma,\epsilon\epsilon}(u,v)\\ \frac{1}{2}F_{+, \Delta, l}^{\sigma\sigma,\epsilon\epsilon}(u,v)& 0)),
\]
and 
\[
      \vec{V}_{\mathrm{identity}} = \mqty(1 & 1) \vec{V}_{+, \Delta=0, l=0} \mqty(1\\1). 
\]
Here, we used the identity $\lambda_{\sigma\sigma\epsilon}=\lambda_{\sigma\epsilon\sigma}$~\cite{Kos:2016ysd}.

We now formulate the dual problem, which casts these crossing equations as an SDP. The (dual) problem is to find a five-component functional $\vec{\alpha}=(\alpha^1, \dots, \alpha^5)$ that satisfies:
\begin{equation}
  \begin{aligned}
  \label{eq:semidefinite_programs}
    \vec{\alpha}&\cdot \vec{V}_{\mathrm{identity}}  = 1, \\
    \vec{\alpha}&\cdot \vec{V}_{+, \Delta, l} \succeq 0, \quad (\Delta \geq d, l=0),\\
    \vec{\alpha}&\cdot \vec{V}_{+, \Delta, l} \succeq 0, \quad (\Delta \geq \Delta_{\mathrm{unitarity}}, l=2, 4, 6,\ldots),\\
    \vec{\alpha}&\cdot \vec{V}_{-, \Delta, l} \geq 0, \quad (\Delta \geq d, l=0),\\
    \vec{\alpha}&\cdot \vec{V}_{-, \Delta, l} \geq 0, \quad (\Delta \geq \Delta_{\mathrm{unitarity}}, l=1,2,3,\ldots),\\
    \vec{\alpha}&\cdot
    \mqty(  
        F_{-, \Delta=\Delta_{\epsilon}, l=0}^{\sigma\sigma,\sigma\sigma}(u,v)\\
        0\\
        F_{-, \Delta=\Delta_{\sigma}, l=0}^{\sigma\epsilon,\sigma\epsilon}(u,v)\\
        F_{-, \Delta=\Delta_{\sigma}, l=0}^{\epsilon\sigma,\sigma\epsilon}(u,v)\\
        -F_{+,\Delta=\Delta_{\sigma}, l=0}^{\epsilon\sigma,\sigma\epsilon}(u,v)
        )\geq 0.
  \end{aligned}
\end{equation}
The existence of such a functional $\vec{\alpha}$ (a dual-feasible solution) implies that the original equations Eq.~\eqref{eq:bootstrap_eq} have no solution for the assumed spectrum, thus ruling out that point in the $(\Delta_\sigma, \Delta_\epsilon)$ parameter space.

To execute this SDP, we used an adapted version of \texttt{simpleboot}~\cite{simpleboot} as a front-end, which uses \texttt{SDPB}~\cite{Landry:2019qug} for solving the semi-definite program \cite{Simmons-Duffin:2015qma}. The cut-off parameter is $\Lambda=19$ and the other numerical details are shown in Appendix A. Whenever comparison is possible, our results presented in this paper are compatible with computations performed using \texttt{cboot} \cite{cboot} and \texttt{PyCFTboot} \cite{Behan:2016dtz}.

\section{The Bounds on $(\Delta_{\sigma}, \Delta_{\epsilon})$}

We now present the numerical bounds derived from the SDP Eq.~\eqref{eq:semidefinite_programs}, which incorporates the duality-inspired fusion rules. We begin with the $d=2$ case.

\begin{figure*}[tb] % Use figure* to span both columns.
    \centering 
    % --- First Subfigure (d=2) ---
    \subfloat[The allowed region in the $(\Delta_{\sigma}, \Delta_{\epsilon})$ plane for $d=2$. \label{fig:2d_bootstrap}]{%
        \includegraphics[width=0.48\textwidth]{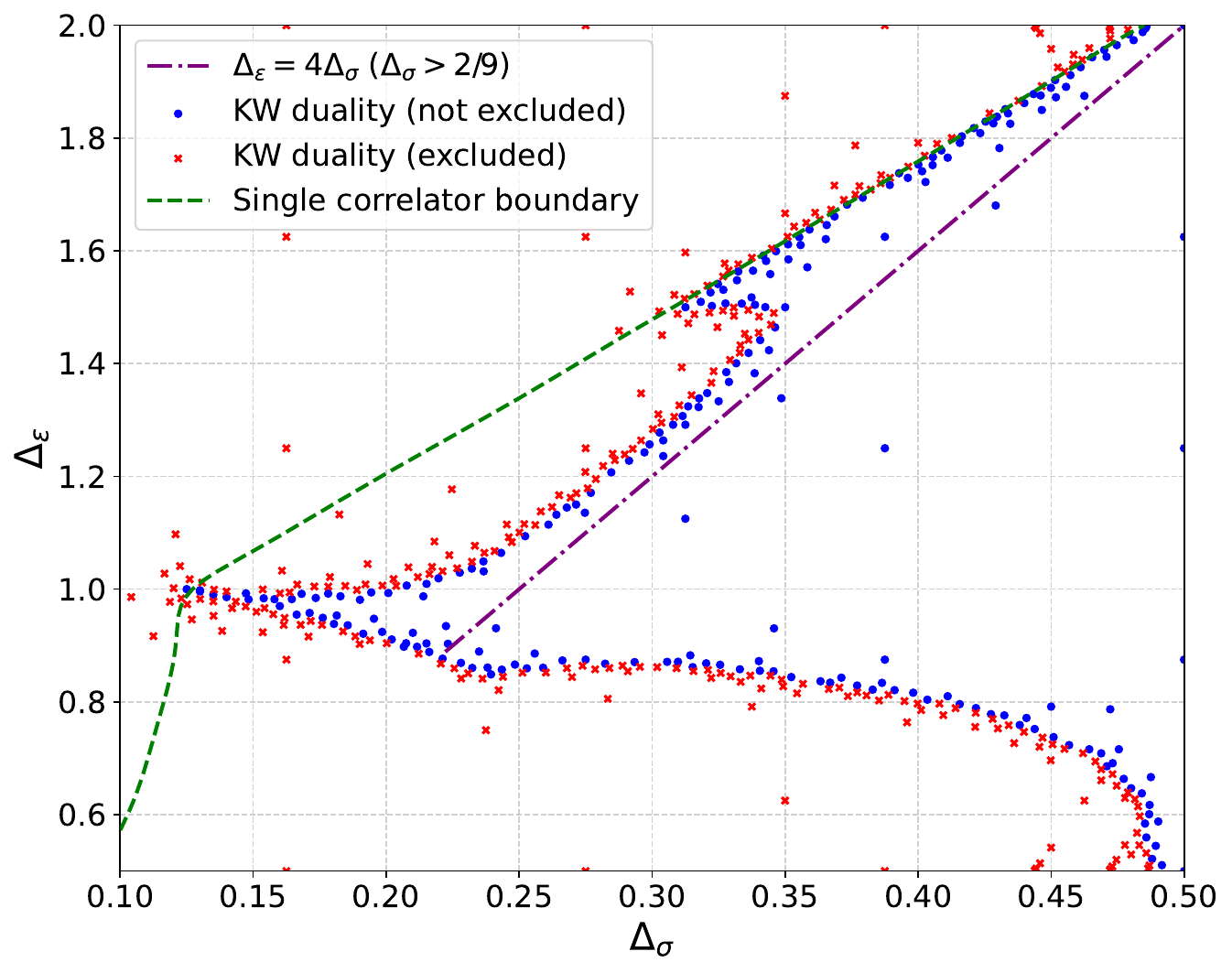}% Adjust width as needed
    }%
    \hfill % Add horizontal space between subfigures
    % --- Second Subfigure (d=3) ---
    \subfloat[The allowed region in the $(\Delta_{\sigma}, \Delta_{\epsilon})$ plane for $d=3$.\label{fig:3d_bootstrap}]{%
        \includegraphics[width=0.48\textwidth]{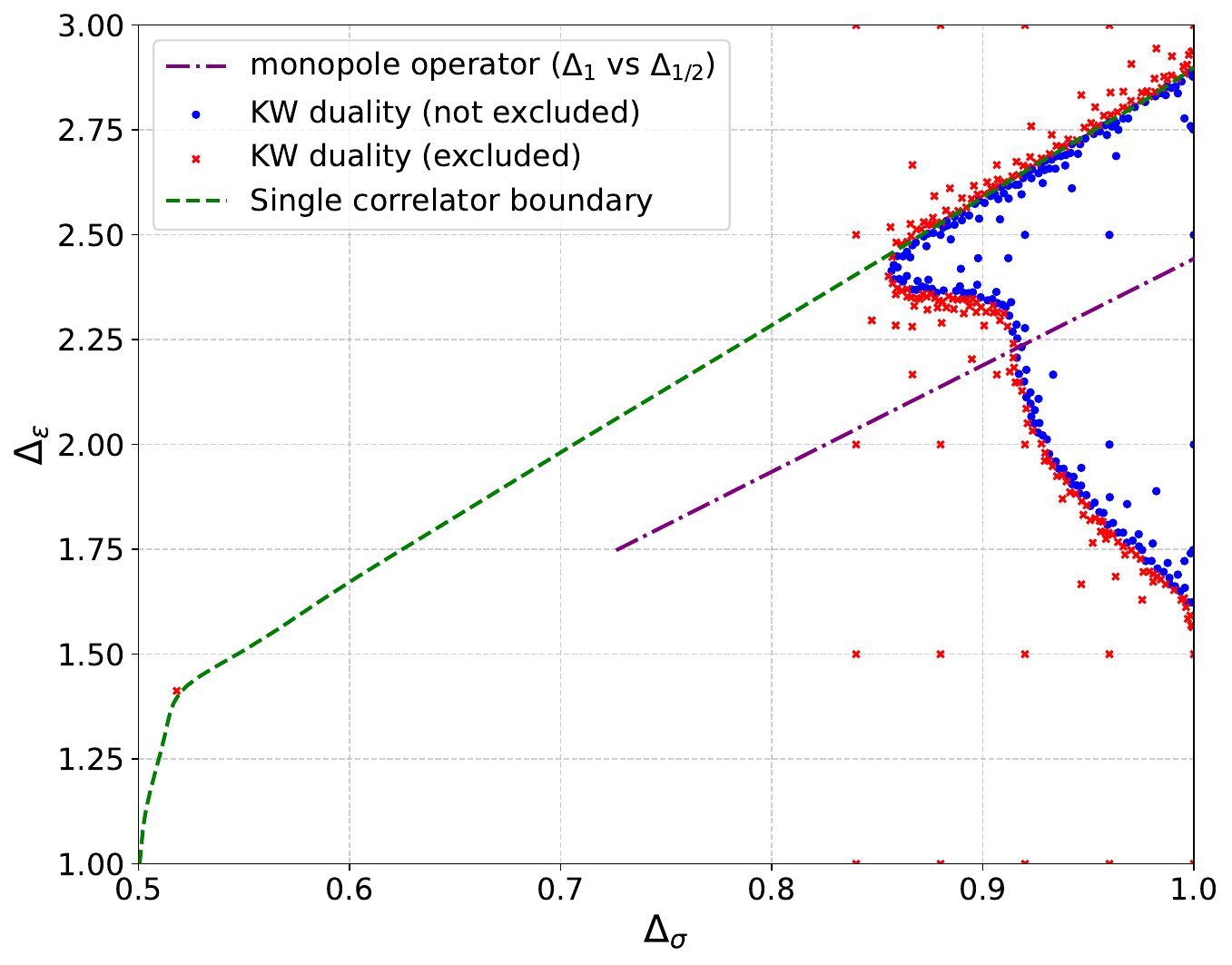}% Adjust width as needed
    }
    % --- Overall caption ---
    \captionsetup{justification=raggedright,singlelinecheck=false}
    \caption{Bootstrap bounds on the $(\Delta_{\sigma}, \Delta_{\epsilon})$ plane imposing the duality-inspired fusion rule for (a) $d=2$ and (b) $d=3$. Blue dots represent the not excluded points, while red crosses are excluded. The purple dash-dotted lines are candidate CFTs discussed in the main text.}
\captionsetup{justification=centering,singlelinecheck=true}
    \label{fig:2d_3d_bootstrap} % Overall label for the figure*
\end{figure*}

Figure~\ref{fig:2d_bootstrap} shows the allowed region in the $(\Delta_{\sigma}, \Delta_{\epsilon})$ plane for $d=2$. We immediately observe two prominent features, which we refer to as the `chin' and the `nose'. 
As expected, the tip of the `chin' coincides precisely with the 2D Ising model, $(\Delta_{\sigma}, \Delta_{\epsilon})=(1/8, 1)$, which is correctly not excluded by our bounds. This is a non-trivial result; we recall that standard $\mathbb{Z}_2$ mixed-correlator bounds in $d=2$ (without our fusion rule) do not isolate the Ising point into an island \cite{Behan:2017rca}, unlike in $d=3$. Thus, our duality-inspired fusion rule provides a significantly stronger constraint near the Ising point.

The second feature is the `nose', whose tip kinks at $(\Delta_{\sigma}, \Delta_{\epsilon})=(5/16, 3/2)$. We conjecture this kink is exact and will not move with increased numerical precision (larger $\Lambda$). This point corresponds precisely to the primary fields $\sigma \equiv \phi_{2,1}$ and $\epsilon \equiv \phi_{3,1}$ in the $\mathcal{M}(8,7)$ minimal model. We find this feature's appearance is critically dependent on the imposed condition $\lambda_{\sigma\sigma \epsilon} = \lambda_{\sigma \epsilon \sigma}$; the `nose' vanishes without it.

This identification is puzzling, as the $\mathcal{M}(8,7)$ model does \textit{not} satisfy our full fusion rules Eq.~\eqref{eq:ising_fusion_rule}. Specifically, its OPE contains $\epsilon \times \epsilon \sim 1 + \epsilon + \cdots$, which explicitly violates our constraint.

Why is this theory not excluded? We observe that at the tip of the `nose' (and for larger $\Delta_{\sigma}$), a phenomenon known as a ``trivial mix" occurs \cite{pirsa_PIRSA:23040146}. When this happens, the components of the functional $\vec{\alpha}$ corresponding to the mixed-correlator equations (i.e., $\alpha^2, \ldots, \alpha^5$) are numerically suppressed to zero, while only the component corresponding to the $\langle\sigma\sigma\sigma\sigma\rangle$ single-correlator (i.e., $\alpha^1$) remains non-zero. The bound effectively reverts to a single-correlator bound, which the $\mathcal{M}(8,7)$ model satisfies \cite{Liendo:2012hy}. It has also been noted that a dual jump does not occur in the region where this happens.

This is not a bug (the bound is still mathematically correct), but it indicates that the power of the \textit{mixed-correlator} constraints is diminished in this region. Our observation that this ``trivial mix" occurs precisely at the location of a known CFT may shed further light on this numerical phenomenon.

Finally, we believe the overall shape of the allowed region reflects the structure of valid CFTs. We will discuss the known CFTs that {\it do} satisfy our duality-inspired fusion rules (corresponding to the purple dashed lines in Fig.~\ref{fig:2d_bootstrap}) in the next section.

We now turn to $d=3$, with results shown in Fig.~\ref{fig:3d_bootstrap}. The first observation is that the `chin' feature has disappeared. Correspondingly, the 3D Ising point, $(\Delta_{\sigma}, \Delta_{\epsilon})=(0.5181489, 1.412625)$~\cite{Kos:2014bka}, is robustly excluded by our bound. This is a key result: the 3D Ising model is known to violate our constraint (its OPE is $\epsilon \times \epsilon \sim 1 + \epsilon + \cdots$). The fact that our duality-inspired bootstrap correctly rules out this theory confirms the effectiveness of the selection rule.

Our bound establishes that any $d=3$ CFT satisfying all constraints can only exist in the region $\Delta_{\sigma} \gtrsim 0.85$. We will discuss the implications of this new bound for three-dimensional gauge theories in the next section.

The lower bound on $\Delta_{\sigma}$ is set by a `nose' feature. As a function of $d$, this `nose' appears continuously connected to the $\mathcal{M}(8,7)$ point discussed for $d=2$ (similar to how the Ising point can be tracked from $d=2$ \cite{El-Showk:2013nia}). While it is tempting to associate this $d=3$ kink with a physical theory, we currently have no immediate candidate. We note that the multi-critical scalar theories, which are a natural interpretation of the $\mathcal{M}(m+1,m)$ series, are not expected to exist in $d=3$.

Our results for dimensions $d=4, 5, 6,$ and $7$ are shown in Appendix B. The comparison with the conventional bound without assuming the duality-inspired fusion rule can be found there as well. As $d$ increases, the allowed region generally expands, and the `nose' feature becomes less pronounced, eventually disappearing into the boundary. This aligns with the general expectation that interacting CFTs might not exist above certain critical dimensions (like $d=4$ or $d=6$), particularly those without gauge fields. We will discuss possible gauge theory candidates compatible with our bounds in the next section.

Finally, we address the possibility of additional, isolated allowed regions (`islands'). In all dimensions studied, we performed extensive scans using lower derivative orders $\Lambda$ —allowing us to probe a wider parameter space—across the ranges shown in the Figures. We found no evidence for such disconnected solutions. While this does not constitute a rigorous proof of uniqueness, it suggests that the connected allowed regions presented here likely represent the complete set of solutions consistent with our assumptions within the explored parameter ranges.

\section{Candidate CFTs Consistent with the Bounds}
\setcounter{equation}{0} % Reset equation counter for this section

In this section, we discuss candidate CFTs that are consistent with the bounds derived in the previous section by satisfying our duality-inspired fusion rules Eq.~\eqref{eq:ising_fusion_rule}.

\textit{Two dimensions.—}
As mentioned in the introduction, CFTs with certain invertible symmetries might satisfy our fusion rules, even without an underlying self-duality. This scenario is indeed realized in two dimensions by free compactified bosons.

Consider a free massless boson $\phi \sim \phi + 2\pi R$ compactified on a circle of radius $R$. The theory possesses a $U(1)$ symmetry. We can identify operators consistent with our fusion rules, for instance, by taking $\sigma \sim \cos(k\phi)$ and $\epsilon \sim \cos(2k\phi)$ for some momentum $k$. These operators have scaling dimensions $\Delta_\sigma \propto k^2$ and $\Delta_\epsilon \propto (2k)^2$, satisfying $\Delta_\epsilon = 4\Delta_\sigma$. The OPEs $\sigma \times \sigma \sim 1 + \epsilon + \dots$ and $\sigma \times \epsilon \sim \sigma + \dots$ hold. Crucially, the $\epsilon \times \epsilon \sim 1 + \cos(4k\phi) + \dots$ OPE does \textit{not} contain $\epsilon \sim \cos(2k\phi)$, thus satisfying our key constraint. Furthermore, for this system to match our assumptions, we require $\sigma$ and $\epsilon$ to be the only relevant scalars, meaning higher harmonics like $\cos(3k\phi)$ (with dimension $9\Delta_\sigma$) must be irrelevant ($\Delta > 2$). The line $\Delta_\epsilon = 4\Delta_\sigma$ corresponding to such free boson theories is plotted in Fig.~\ref{fig:2d_bootstrap} (purple dash-dotted line). It remains an interesting question whether features of our bound, especially upon increasing $\Lambda$, relate to specific points or radii along this line.

As for non-candidates, we note that the tricritical Ising model, despite possessing KW duality, is excluded by our setup. This is because it is tricritical, meaning it has more than two relevant scalar operators ($\sigma$ and $\epsilon$), violating our initial assumption designed to isolate the minimal fine-tuning scenario. Our bounds correctly reflect this, lying away from the tricritical Ising point.

\textit{Three dimensions.—}
In three dimensions, while self-duality is known to occur in some supersymmetric gauge theories (e.g., via mirror symmetry \cite{Gaiotto:2008ak,Cremonesi:2013lqa}), it is not immediately clear if these dualities manifest as the specific fusion rule we impose on low-lying scalar operators. Instead, we focus on theories with $U(1)$ symmetry (or its discrete subgroups) as potential realizations to test our bounds.

An interesting candidate is QED$_{3}$ with $N_f$ flavors of massless Dirac fermions, which flows to an interacting CFT for sufficiently large $N_f$ 
and has been extensively studied using the conformal bootstrap \cite{Chester:2016wrc, He:2021sto, Albayrak:2021xtd}.
This theory possesses monopole operators $\mathcal{M}_q$ carrying monopole charge $q$ under a topological $U(1)_J$ symmetry. 
We define the operators corresponding to the lowest and next-lowest topological charges as $\sigma \equiv O_{1/2}$ and $\epsilon \equiv O_{1}$. Schematically, they take the form $O_q \sim \mathcal{M}_q + (\mathcal{M}_{q})^\dagger$. However, to obtain real, parity-definite scalars that precisely satisfy our fusion rules, we must take specific linear combinations of the $SU(N_f)$ components of the monopole operators. With this choice, their OPEs take the required form: $\sigma \times \sigma \sim 1 + \epsilon + \cdots$, $\epsilon \times \epsilon \sim 1 + O_2 + \cdots$, and $\sigma \times \epsilon \sim \sigma + O_{3/2} + \cdots$.

Our bootstrap constraints apply, provided that $\sigma$ and $\epsilon$ are the only relevant scalars appearing in the OPEs, i.e., operators like $O_{3/2}$, $O_2$, and any non-identity $U(1)_J$-singlet scalars are irrelevant ($\Delta > 3$).
The absence of relevant singlet scalars is motivated by the idea that four-fermi operators must be irrelevant in the conformal QED$_3$, making it an example of self-organized criticality (which motivates interest in this model as a description of quantum spin liquids in condensed matter physics \cite{2004PhRvB..70u4437H}). 
Furthermore, in three dimensions, scalar fermion bilinears are parity-odd and thus do not appear in the OPE if $\sigma$ and $\epsilon$ are parity-definite.

More precisely, the monopole operator $M_{+1/2}$ with topological charge $q=1/2$ is dressed by $N_f/2$ fermion zero modes. Consequently, it is scalar (spin-zero) under $SO(3)$ rotations and belongs to the rank-$N_f/2$ antisymmetric tensor representation of the $SU(N_f)$ flavor symmetry~\cite{Borokhov:2002ib}. 
The Hermitian conjugate of the monopole operator is given by $(M_{+1/2}^{a_1 \dots a_{N_f/2}})^\dagger = (M_{-1/2})_{a_1 \dots a_{N_f/2}}$. Under a parity transformation $P$, it transforms as
$PM_{+1/2}^{a_1\cdots a_{N_f/2}}P^{-1} = \eta M_{-1/2}^{a_1\cdots a_{N_f/2}},$
where $\eta$ is a phase factor. 
To construct a real, parity-definite scalar operator $\sigma$, we consider a linear, manifestly real combination of specific $SU(N_f)$ components:
\begin{equation*}
    \begin{aligned}
        \sigma &= \alpha M_{+1/2}^{1 \dots N_f/2} + \alpha^* (M_{-1/2})_{1 \dots N_f/2} \\ 
        &\quad + \beta M_{+1/2}^{N_f/2+1 \dots N_f} + \beta^* (M_{-1/2})_{N_f/2+1 \dots N_f}.
    \end{aligned}
\end{equation*}
Requiring this operator to be parity-definite yields the condition $\alpha = (-1)^{N_f/2} \alpha$. Therefore, for $N_f \in 4\mathbb{Z}_{>0}$, we can choose non-zero coefficients to construct a non-trivial real operator with fixed parity. However, for $N_f \in 4\mathbb{Z}_{>0} + 2$, the condition forces $\alpha = \beta = 0$, making it impossible to construct a $\sigma$ that is simultaneously real and parity-definite. We define $\epsilon$ as a real, parity-definite linear combination of the lowest-dimension $|q|=1$ monopole operators appearing in the $\sigma\times\sigma$ OPE. These operators belong to the $SU(N_f)$ representation characterized by a Young diagram with two columns and $N_f/2$ rows. From representation theory, the specific $SU(N_f)$ representation corresponding to $\epsilon$ appears with a multiplicity of exactly one. This guarantees that $\epsilon$ appears exactly once in the $\sigma \times \sigma$ OPE, and similarly, $\sigma$ appears exactly once in the $\sigma \times \epsilon$ OPE. Furthermore, no other operators carrying different $SU(N_f)$ labels appear in these specific fusion channels. These ensure that our assumption $\lambda_{\sigma\sigma \epsilon} = \lambda_{\sigma \epsilon \sigma}$ is valid. In addition, we assume that all other scalar operators in both the $|q|=1/2$ and $|q|=1$ sectors are irrelevant.

The scaling dimensions of monopole operators in QED$_3$ have been computed using $1/N_f$ expansions \cite{dyer2013monopoletaxonomythreedimensionalconformal}. The purple dash-dotted line in Fig.~\ref{fig:3d_bootstrap} represents the leading non-trivial order prediction relating $\Delta_\sigma = \Delta(O_{1/2})$ and $\Delta_\epsilon = \Delta(O_1)$. This line intersects our allowed region at $\Delta_{\sigma}^* \approx 0.92$. Using the $1/N_f$ results, this corresponds to $N_f^* \approx 3.6$. While this provides an intriguing connection, we reiterate that applying the bootstrap to the full $U(1)_J$ symmetry (if possible) might yield stronger constraints than our current approach based only on the specific fusion rule.

\textit{Four dimensions.—}
Turning to $d=4$, we consider conformal gauge theories as potential candidates. These include non-Abelian gauge theories coupled to fermionic matter in specific representations, residing within the conformal window (e.g., Banks-Zaks-like theories; see \cite{Aarts:2023vsf} and references therein for lattice studies). Assuming such a theory possesses a $U(1)$ flavor symmetry (or a discrete $\mathbb{Z}_n$ subgroup thereof, $n\ge 6$), we can identify the lowest-dimension scalar operators transforming with charges $q=1$ and $q=2$. In many candidate theories, these correspond schematically to a fermion bilinear $\sigma \sim {\psi}\psi + \bar{\psi} \bar{\psi}$ (charge 1) and a four-fermion operator $\epsilon \sim  (\psi \psi)^2 + (\bar{\psi} \bar{\psi})^2$ (charge 2).

If these operators, $\sigma$ and $\epsilon$, are the only relevant scalars, then our bound applies. Our numerical result, $\Delta_{\sigma} \gtrsim 1.4$ for $d=4$ (see Fig.~3(a) in Appendix B), provides a universal lower bound on the dimension of the charge-1 scalar operator in any such theory. The crucial assumption remains the absence of relevant $U(1)$-singlet scalar operators other than the identity. The validity of this assumption depends on the specific theory; for a theory to be conformal without fine-tuning, such operators must be irrelevant. As in $d=3$, for specific theories possessing larger flavor symmetries, imposing the constraints from the full symmetry group in the bootstrap analysis would likely yield stronger bounds than our analysis.

\section{Discussion}
In this paper, we have opened up a new paradigm to constrain the conformal data not from the invertible symmetries but from the fusion rule inspired by the categorical symmetries. While we have focused on the scalar correlation functions, an important future direction is to extend this analysis to operators with Lorentz spin, particularly anti-symmetric tensor fields. Applying similar duality-inspired constraints to their correlation functions could provide non-perturbative bounds relevant to electromagnetic duality in $d=4$ and (self-dual) tensor theories via S-duality in $d=6$.

Our methodology also connects with recent ideas in high-energy physics where non-invertible symmetries are explored as constraints on the structure of quantum field theories and their parameters \cite{Kaidi:2024wio,Cordova:2024vsq, Kobayashi:2024yqq, Bharadwaj:2024gpj, Hidaka:2024kfx, Kobayashi:2025znw, Suzuki:2025oov, Dong:2025pah, Liang:2025dkm, Dong:2025jra, Suzuki:2025bxg,Suzuki:2025kxz}. This work provides a concrete non-perturbative technique, via the conformal bootstrap, to probe the consequences of such symmetries directly at the level of CFT data.

The landscape of CFTs constrained by non-invertible symmetries remains largely unexplored, as a complete classification of such symmetries is still lacking. Applying the conformal bootstrap with selection rules motivated by these generalized symmetries, as demonstrated here, offers a powerful tool for discovering new CFTs and understanding the fundamental principles governing quantum field theory beyond the paradigm of conventional symmetries. Exploring CFTs with fusion rules inspired by such generalized symmetries represents a blue ocean full of potential to attack unsolved problems in theoretical physics.

%More non-trivial fusion rules from categorical symmetry.

\section{Acknowledgements}
We would like to thank C.~Behan and N.~Su for helping us with the numerics via email exchange. YN is in part supported by JSPS KAKENHI Grant Number 21K03581.

\bibliography{main}

\appendix
\section{Numerical parameters}

The parameters for generating the conformal block tables in \texttt{simpleboot} are:
\begin{center}
\texttt{"dim"->d},\texttt{"$\lambda$max"->19},
\texttt{"$\kappa$"->14}, \texttt{"rN"->56},\texttt{"lset"->}\texttt{Range[0,26]}\texttt{\textasciitilde Join\textasciitilde\{49,52\}} 
\end{center}
This parameter set is the ``canonical one" used first in \cite{Kos:2016ysd} for the mixed correlator bootstrap in the 3D Ising model.

Furthermore, the parameters for SDPB are:
\begin{center}
  \texttt{--maxIterations=1000} \texttt{--dualityGapThreshold=1e-50}
  \texttt{--primalErrorThreshold=1e-60} \texttt{--dualErrorThreshold=1e-10}
  \texttt{--precision=768} \texttt{--initialMatrixScalePrimal=1e+20} 
  \texttt{--initialMatrixScaleDual=1e+20} \texttt{--maxComplementarity=1e+70} 
  \texttt{--findPrimalFeasible} \texttt{--findDualFeasible}  \texttt{--detectPrimalFeasibleJump} 
  \texttt{--detectDualFeasibleJump}
\end{center}
Here, in addition to \texttt{--detectPrimal(Dual)FeasibleJump}, we added \texttt{--findPrimal(Dual)Feasible} and also disabled the hot-start feature. It would be more appropriate (and sometimes faster) if we did not use \texttt{--findPrimal(Dual)\allowbreak Feasible}, but 
for reasons mentioned in the main text, our mixed correlator conformal bootstrap did not work in certain regions without these options.

Numerically, we observe that achieving convergence and precise bounds requires larger derivative orders ($\Lambda$) as $d$ increases, a common feature in numerical bootstrap studies. While the location of the `nose' in $d=2$ was quite stable with $\Lambda$, its position and prominence become more sensitive to $\Lambda$ in higher dimensions, further suggesting it might be an artifact related to specific low-dimensional theories rather than a generic feature across all $d$.

\section{More plots in different dimensions}

\begin{figure*}[htb]
    % --- 1行目 (2枚) ---
    \subfloat[$d=2$\label{fig:sub-1}]{%
        \includegraphics[width=0.30\textwidth]{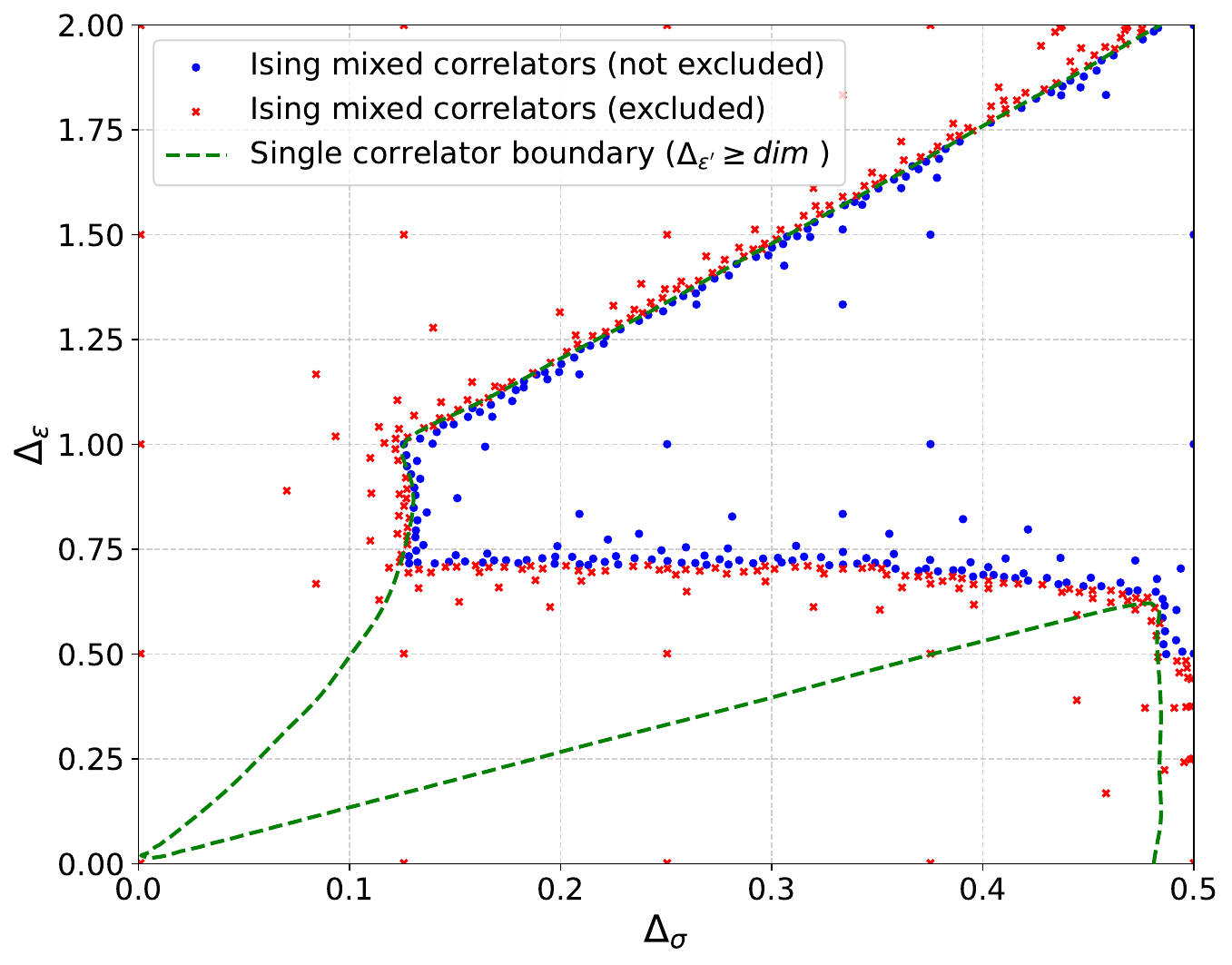}%
    }%
    \hfill 
    \subfloat[$d=3$\label{fig:sub-2}]{%
        \includegraphics[width=0.30\textwidth]{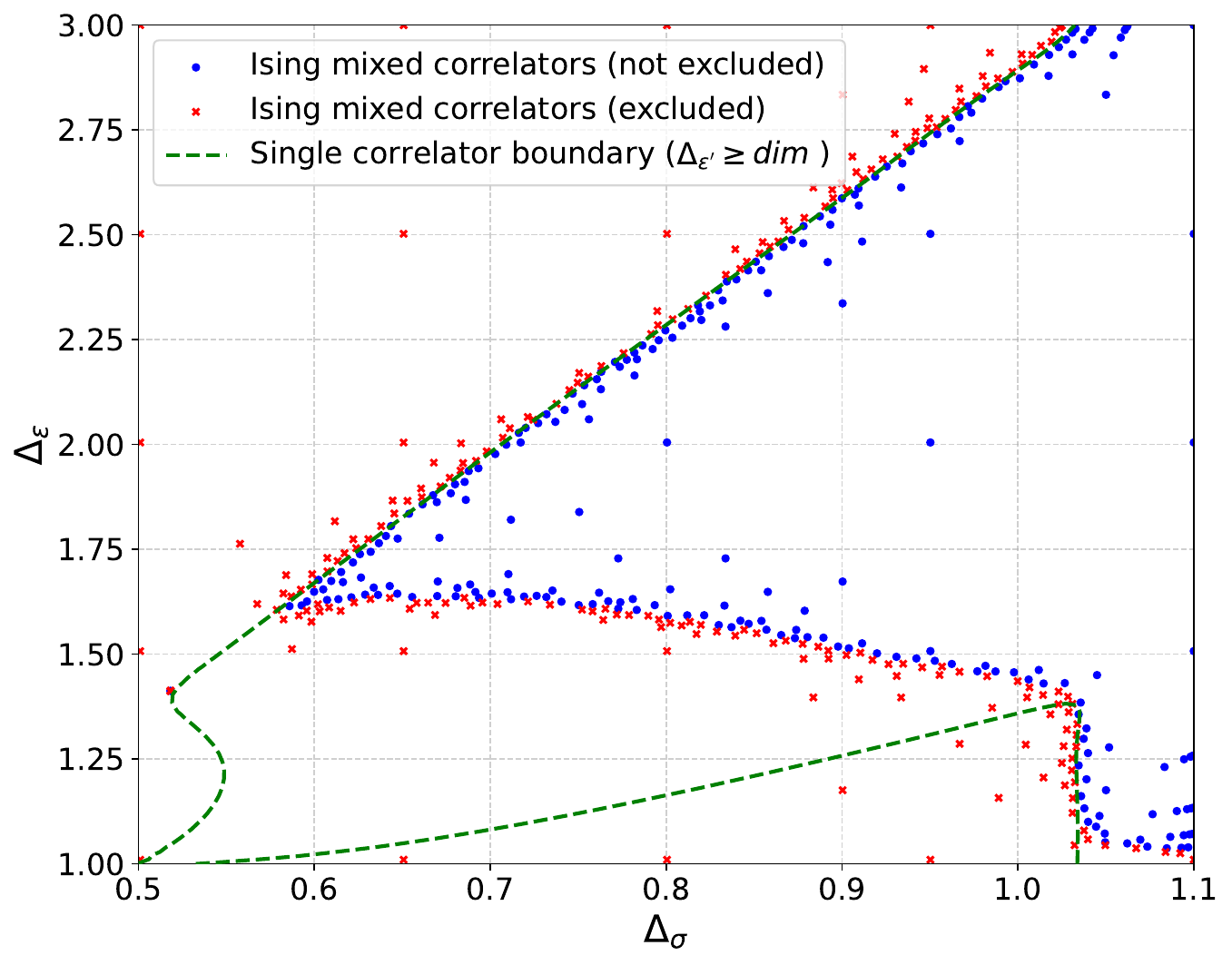}%
    }
    \hfill 
    \subfloat[$d=4$\label{fig:sub-2}]{%
        \includegraphics[width=0.30\textwidth]{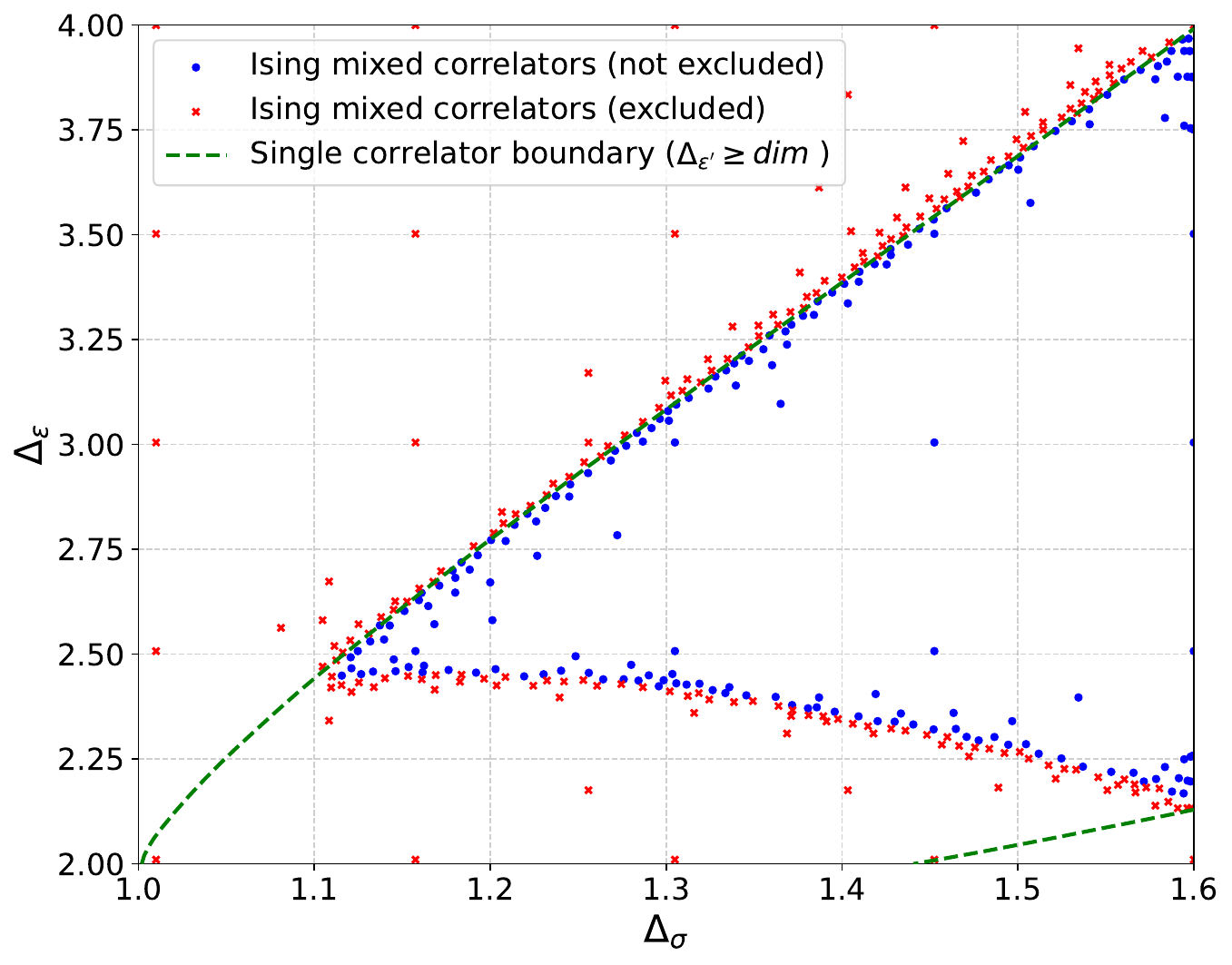}%
    }
    \\ % 改行
    % --- 2行目 (1枚) ---
    \subfloat[$d=5$\label{fig:sub-3}]{%
        \includegraphics[width=0.30\textwidth]{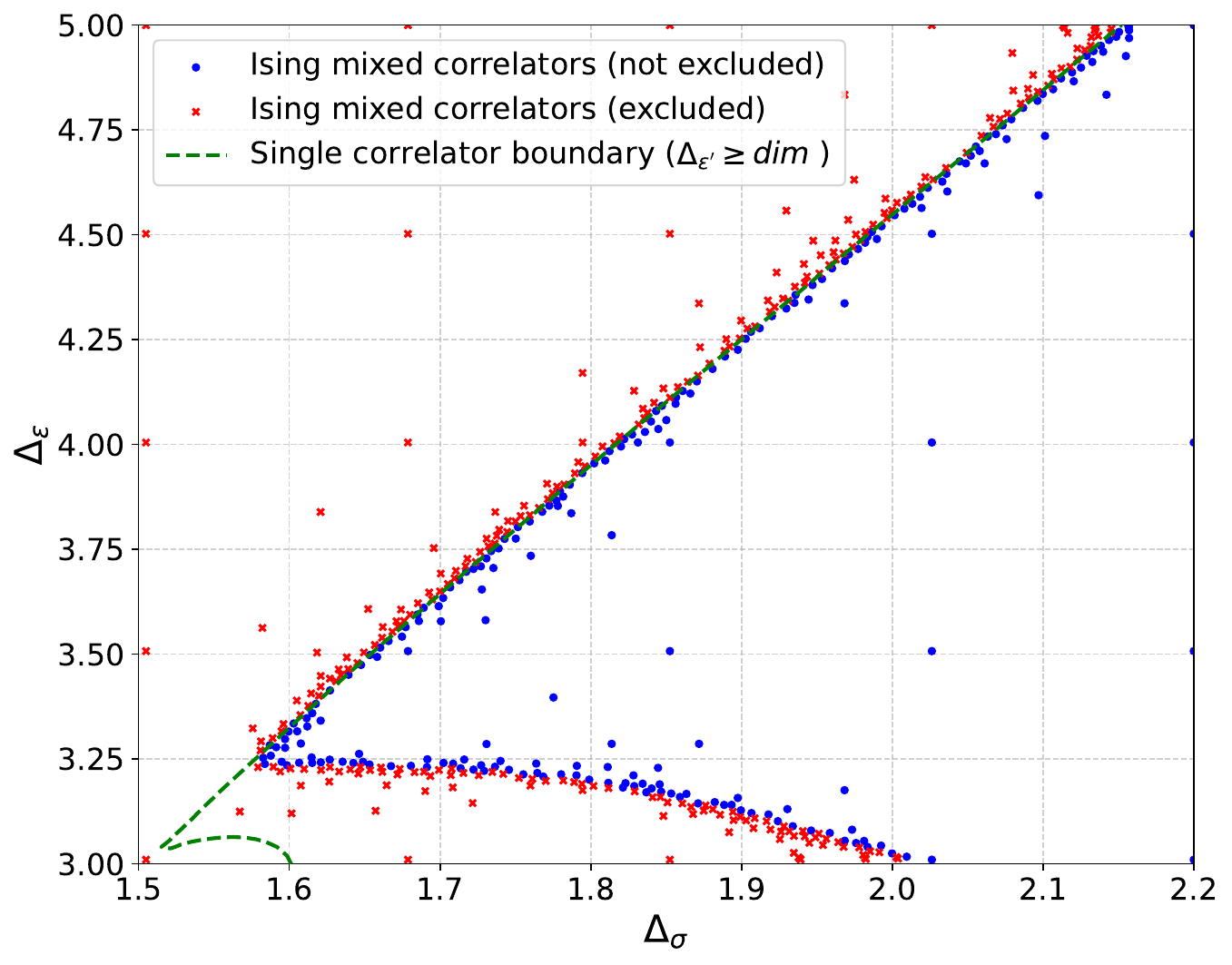}%
    }
    % \hfill 
    \hspace{2em}
    \subfloat[$d=6$\label{fig:sub-4}]{%
        \includegraphics[width=0.30\textwidth]{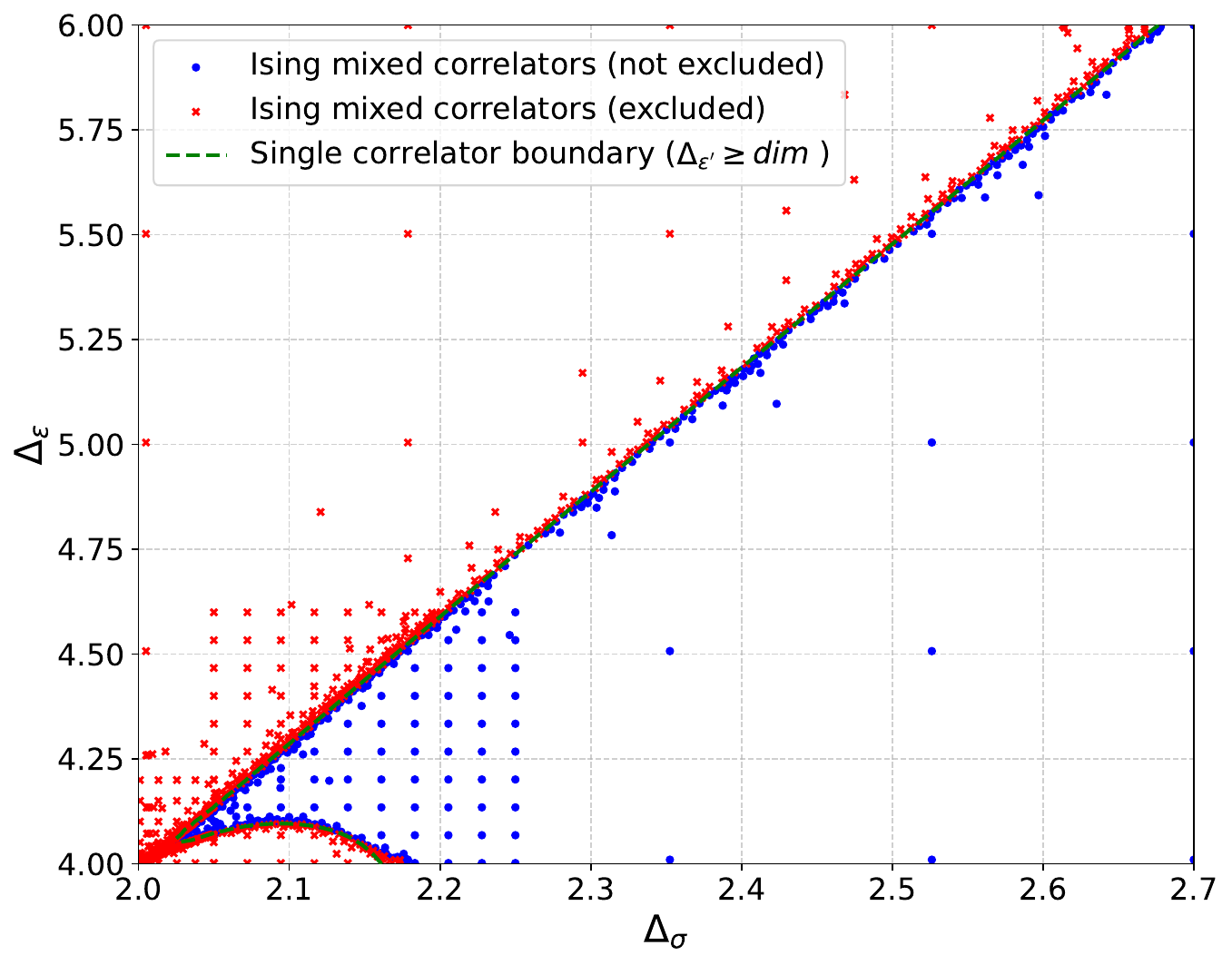}%
    }
    % --- Overall caption (optional) ---
    \caption{Bootstrap bounds on the $(\Delta_{\sigma}, \Delta_{\epsilon})$ plane imposing the conventional Ising bootstrap constraints. The single correlator bound imposes a gap on $\Delta_{\epsilon'}$.  Blue dots represent the not excluded points, while red crosses are excluded.}
    \label{fig:higher_dim}
\end{figure*}

In this Appendix, we present additional plots to demonstrate the effectiveness of the duality-inspired conformal bootstrap. Figure~\ref{fig:higher_dim} displays the bootstrap bounds for $\mathbb{Z}_2$-symmetric (Ising-type) CFTs derived from the mixed system of correlation functions involving $\sigma$ and $\epsilon$: $\langle \sigma\sigma\sigma\sigma \rangle$, $\langle \epsilon\epsilon\epsilon\epsilon \rangle$, and $\langle \sigma\sigma\epsilon\epsilon \rangle$, across various dimensions. To facilitate comparison with our duality-inspired results in the main text, we impose the gap assumptions $\Delta_{\epsilon'} \ge d$ and $\Delta_{\sigma'} \ge d$, along with the OPE constraint $\lambda_{\sigma \sigma \epsilon} = \lambda_{\sigma \epsilon \sigma}$.

As observed in \cite{Behan:2016dtz,Behan:2017rca}, the assumption $\lambda_{\sigma \sigma \epsilon} = \lambda_{\sigma \epsilon \sigma}$ is crucial for obtaining a bound in $d=2$ that improves upon the single-correlator bootstrap. In $d=3$, this assumption results in an allowed `peninsula' that is smaller than the one reported in \cite{Kos:2014bka}. To our knowledge, the results for $d\ge 4$ are new. We observe that the allowed `peninsula' touches the free scalar point $(\Delta_\sigma, \Delta_{\epsilon}) = (\frac{d-2}{2}, d-2)$ for $d \ge 6$. This is because the condition $\Delta_{\sigma'} \ge d$ becomes compatible with the free scalar spectrum in this range.

The plots indicate a necessary condition $\Delta_{\sigma} \ge \Delta_*$ to realize (self-organized) criticality without fine-tuning under these assumptions: $\Delta_* \approx 0.5, 1.03, 1.61, 2.16$, and $2.68$ in $d=2, 3, 4, 5$, and $6$ dimensions, respectively. These bounds could likely be tightened by increasing the derivative order $\Lambda$, which is expected to be more pronounced in higher dimensions.

\begin{figure*}[htb]
    % --- 1行目 (2枚) ---
    \subfloat[$d=4$\label{fig:sub-1}]{%
        \includegraphics[width=0.33\textwidth]{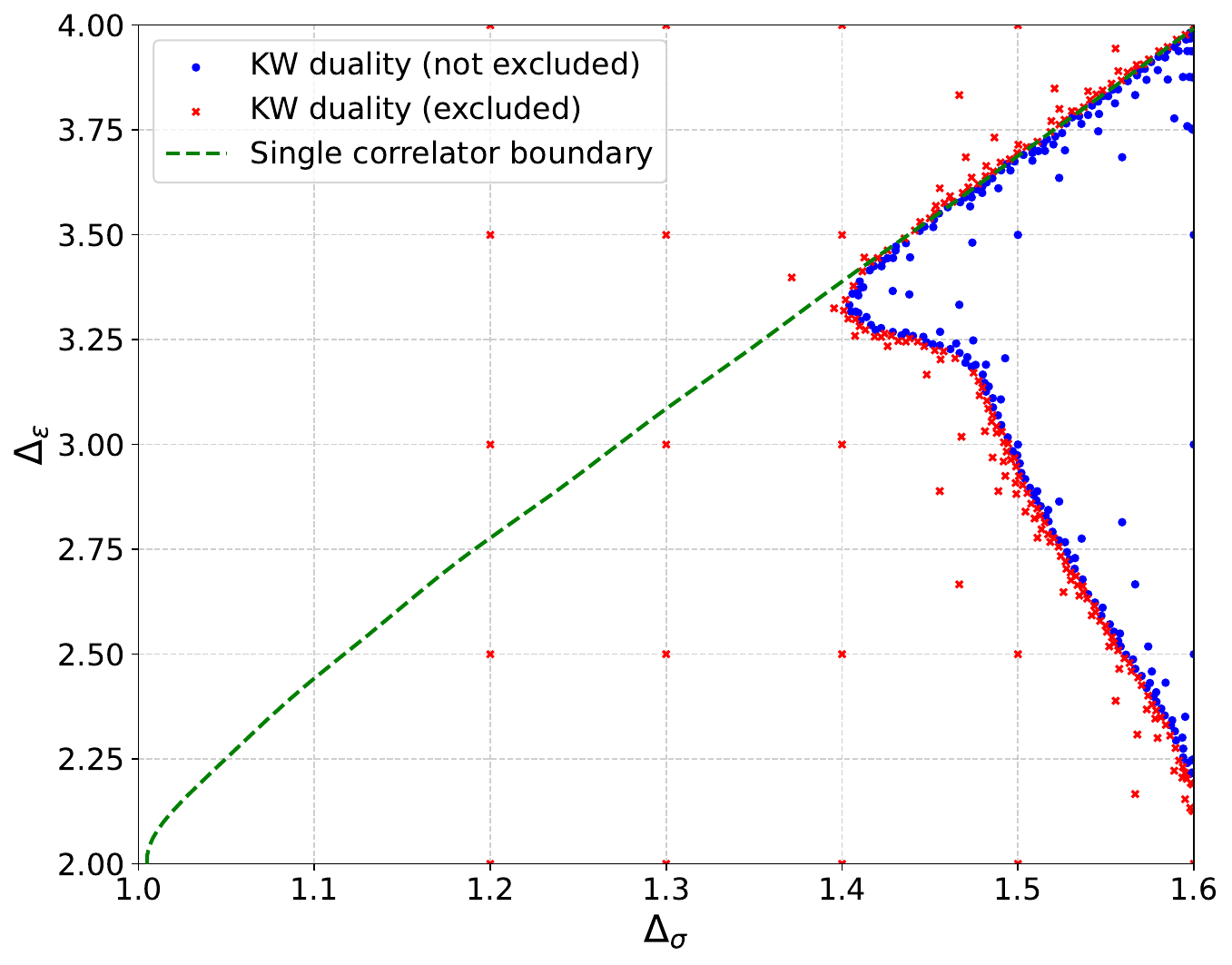}%
    }%
    % \hfill 
    \hspace{2em}
    \subfloat[$d=5$\label{fig:sub-2}]{%
        \includegraphics[width=0.33\textwidth]{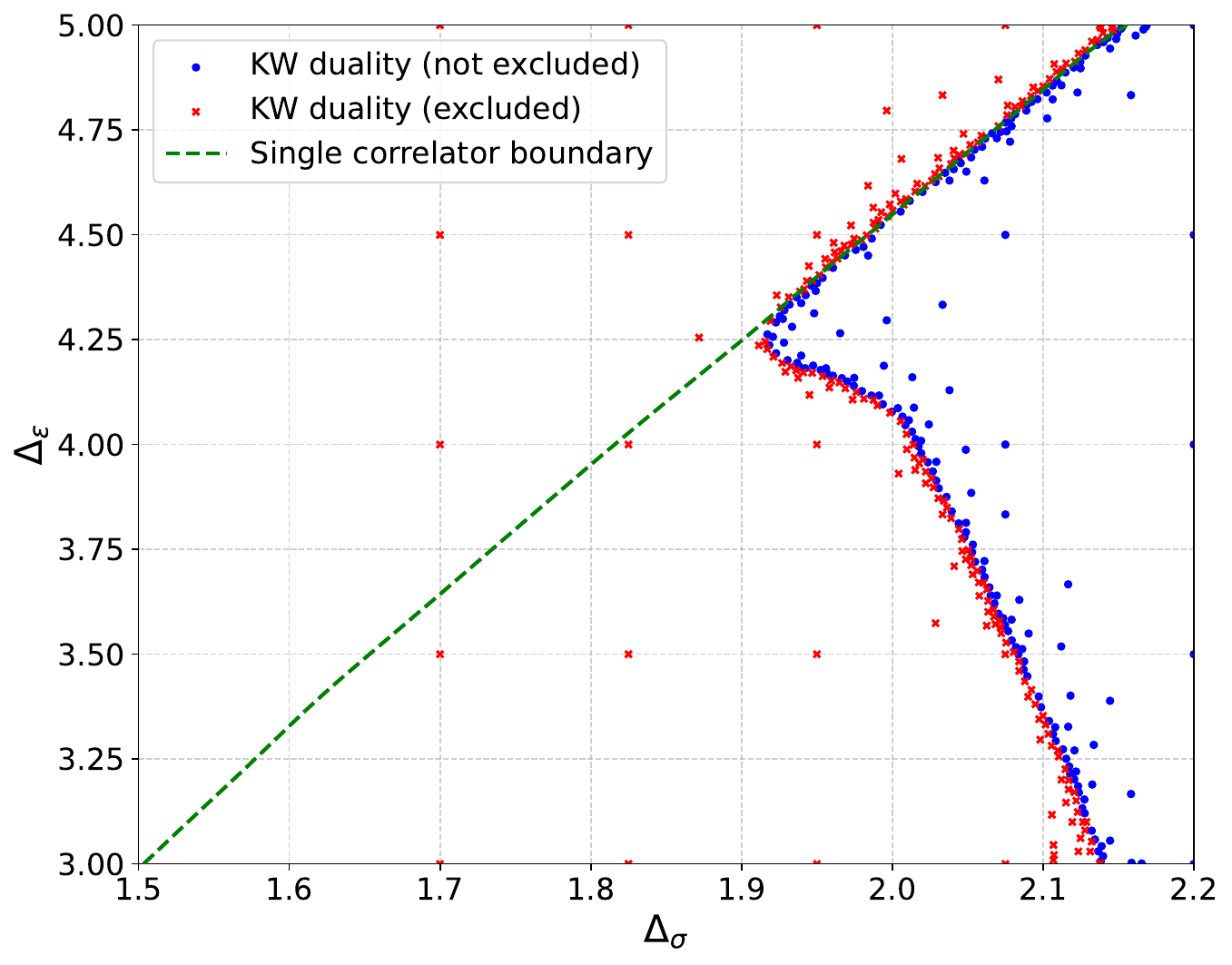}%
    }
    \\ % 改行
    % --- 2行目 (1枚) ---
    \subfloat[$d=6$\label{fig:sub-3}]{%
        \includegraphics[width=0.33\textwidth]{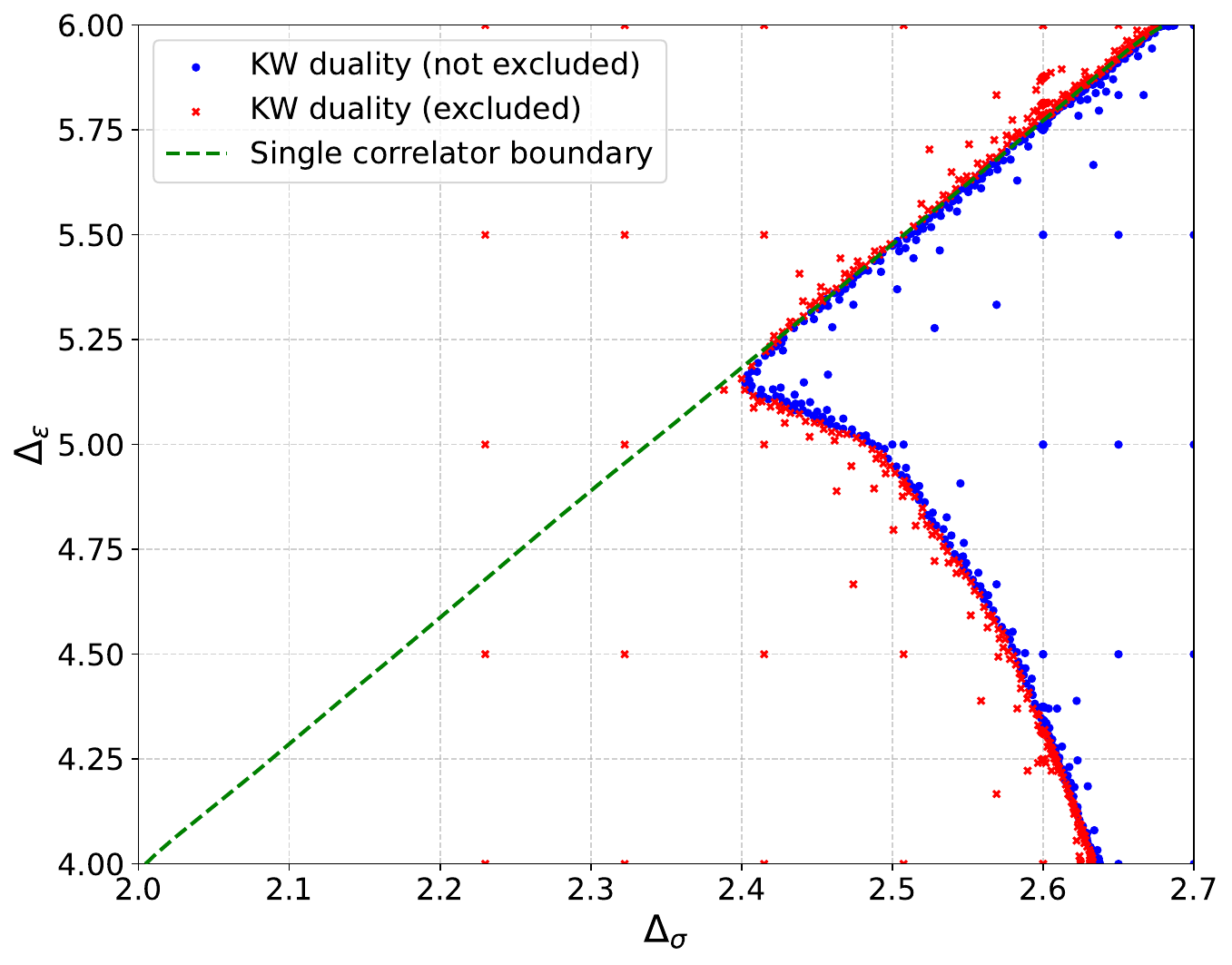}%
    }
    % \hfill 
    \hspace{2em}
    \subfloat[$d=7$\label{fig:sub-4}]{%
        \includegraphics[width=0.33\textwidth]{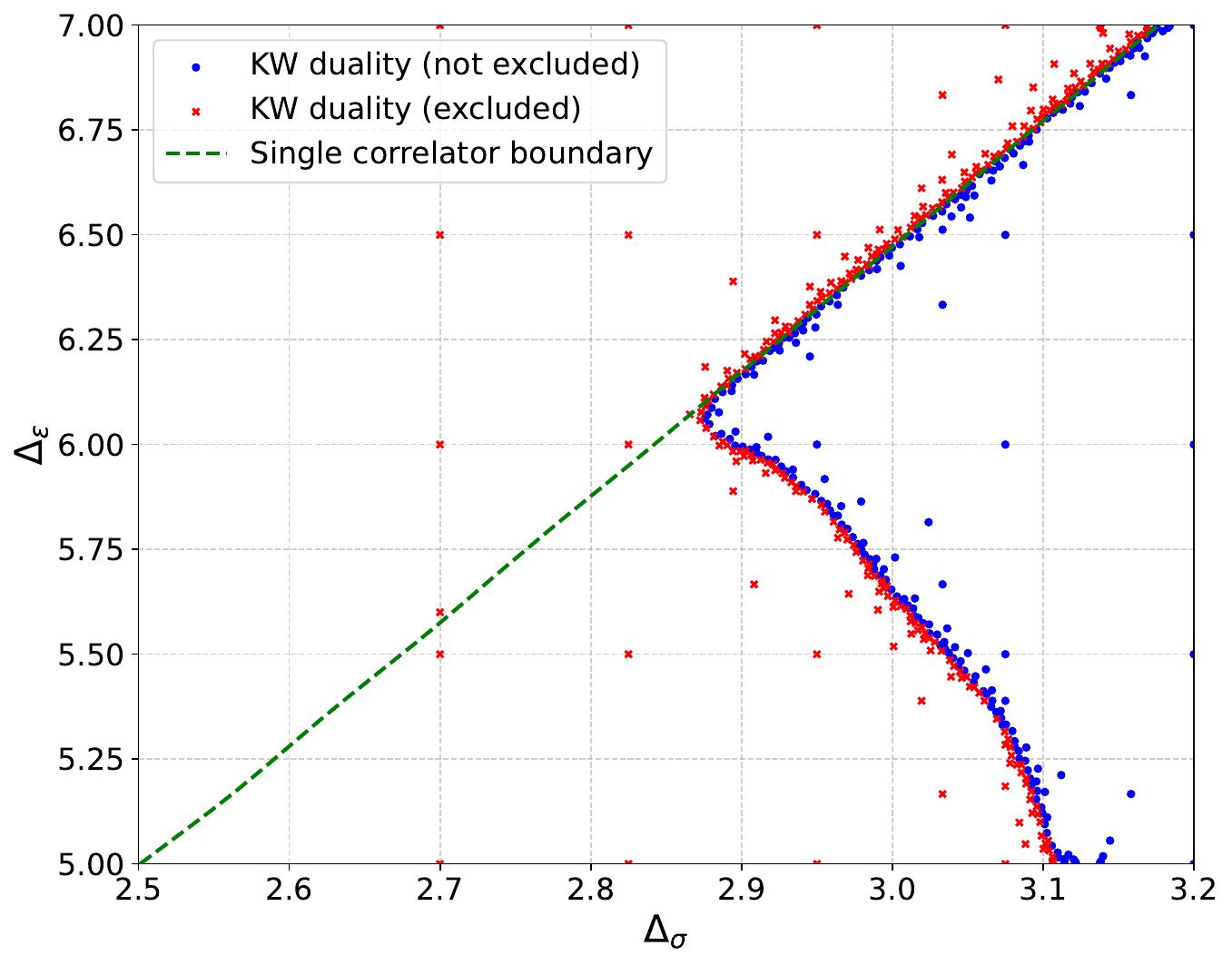}%
    }
    % --- Overall caption (optional) ---
    \caption{Bootstrap bounds on the $(\Delta_{\sigma}, \Delta_{\epsilon})$ plane imposing the duality-inspired fusion rule for higher dimensions: (a) $d=4$, (b) $d=5$, (c) $d=6$, and (d) $d=7$. Blue dots represent the not excluded points, while red crosses are excluded.}
    \label{fig:higher_dim_KW}
\end{figure*}

We now present the plots for the duality-inspired conformal bootstrap in higher dimensions. While the results for $d=2$ and $d=3$ are displayed in the main text, we provide the corresponding bounds for $d=4, 5, 6,$ and $7$ in this document. A comparison between Fig.~\ref{fig:higher_dim} and Fig.~\ref{fig:higher_dim_KW} demonstrates the significantly stronger constraints on the allowed conformal dimensions imposed by the duality-inspired bootstrap.

With the duality-inspired OPE imposed, the necessary condition $\Delta_{\sigma} \ge \Delta_*$ for realizing (self-organized) criticality without fine-tuning is modified to: $\Delta_* \approx 0.125, 0.85, 1.4, 1.92, 2.4$, and $2.88$ in $d=2, 3, 4, 5, 6$, and $7$ dimensions, respectively. As before, these values could be improved by increasing $\Lambda$. These results suggest that categorical symmetry can significantly expand the possibility of finding criticality without fine-tuning (by allowing regions that might be excluded without the duality assumptions).

\end{document}